\documentclass[12pt,a4paper]{iopart}
\usepackage{graphicx}
\begin{document}
\title{The Tsallis Distribution 
in Proton-Proton Collisions at $\sqrt{s}$ = 0.9 TeV at the LHC.}
\author{J. Cleymans and D. Worku}
\address{UCT-CERN Research Centre and Department of Physics, University of Cape Town, 
Rondebosch 7701, South Africa}
\date{\today}
%
%
\begin{abstract}
The Tsallis distribution has been used recently to fit the 
transverse momentum distributions of identified particles
by  the STAR~\cite{Abelev:2006cs} and PHENIX~\cite{Adare} 
collaborations at 
the Relativistic Heavy Ion Collider 
and by the ALICE~\cite{Aamodt:2011zj} and CMS~\cite{Khachatryan:2011tm} 
collaborations at the Large Hadron Collider.
Theoretical issues are clarified concerning the thermodynamic 
consistency of the Tsallis distribution 
in the particular case of 
relativistic high energy quantum distributions.    
An improved  form is proposed for describing the transverse 
momentum distribution and fits are presented together with estimates of
the parameter $q$ and the temperature $T$.
\end{abstract}
%
%
\section{\label{secIntroduction}Introduction}
The Tsallis distribution has gained prominence recently  
in high energy physics with very high quality fits
of the transverse momentum distributions  made
by the STAR~\cite{Abelev:2006cs} and PHENIX~\cite{Adare} collaborations at the 
Relativistic Heavy Ion Collider 
and by the ALICE~\cite{Aamodt:2011zj} and CMS~\cite{Khachatryan:2011tm}
 collaborations at the Large Hadron Collider.

In the literature there exists more than one version of the Tsallis 
distribution~\cite{Tsallis:1987eu,Tsallis:1998ab}
and we investigate in this paper  a version which 
 we consider suited for
describing results in high energy particle physics.
Our main guiding criterium will be thermodynamic consistency which
has not always been implemented 
correctly (see e.g.~\cite{Pereira:2007hp,Pereira:2009ja,Conroy:2010wt}).
The explicit form which we use  is: 
\begin{eqnarray}
&&\frac{dN}{dp_T~dy} = \nonumber\\
&&gV\frac{p_Tm_T\cosh y}{(2\pi)^2}
\left[1+(q-1)\frac{m_T\cosh y -\mu}{T}\right]^{q/(1-q)},
\label{Tsallis-B}
\end{eqnarray}
where $p_T$ and $m_T$ are the transverse momentum and mass respectively, $y$
is the rapidity, $T$ and $\mu$ are the temperature and the chemical potential,
$V$ is the volume, $g$ is the degeneracy factor.
In the limit where the parameter $q$ goes to 1 this reproduces 
the standard Boltzmann distribution:
\begin{eqnarray}
&&\lim_{q\rightarrow 1}\frac{dN}{dp_T~dy} = \nonumber\\
&&gV\frac{p_Tm_T\cosh y}{(2\pi)^2}
\exp\left(-\frac{m_T\cosh y -\mu}{T}\right).
\label{boltzmann}
\end{eqnarray}
In order to distinguish Eq.~\ref{Tsallis-B} from the form used by 
the STAR, PHENIX, ALICE and CMS 
collaborations~\cite{Abelev:2006cs,Adare,Aamodt:2011zj, Khachatryan:2011tm}. 
The motivation for preferring this form is presented in detail 
in the rest of this paper.
The parametrization given in Eq.~(\ref{Tsallis-B}) is close (but different) from
the one used by STAR, PHENIX, ALICE and CMS:
\begin{equation}
  \frac{{\rm d}^2N}{{\rm d}p_{\rm T}{\rm d}y} = p_{\rm T} \frac{{\rm d}N}
  {{\rm d}y} \frac{(n-1)(n-2)}{nC(nC + m_{0} (n-2))} \left( 1 + \frac{m_{\rm T} - m_{0}}{nC} \right)^{-n}
\label{ALICE-CMS}
\end{equation}
where $n$, $C$ and $m_0$ are fit parameters.
The analytic expression used in Refs.~\cite{Abelev:2006cs,Adare,Aamodt:2011zj,Khachatryan:2011tm} 
corresponds to identifying 
\begin{equation}
n\rightarrow \frac{q}{q-1}
\end{equation}
and 
\begin{equation}
nC  \rightarrow \frac{T}{q-1}
\end{equation}
But differences do not allow for the above identifiction to be made 
complete due to an additional factor of the transverse mass on
the right-hand side and
a shift in the tansverse mass.
They are close but not the same.
In particular, no clear pattern emerges for the values of $n$ and $C$ while
an interesting regularity is obtained for $q$ and $T$ as seen in 
Figs.~(\ref{q_value}) and (\ref{T_value}) shown at the end of this paper.
\\
In the next section we review the derivation of the 
Tsallis distribution by emphasizing the quantum statistical form and  the
thermodynamic consistency. 
%
%
%
\section{Tsallis Distribution for Particle Multiplicities.}
%
In the following we discuss the Tsallis form of  the Fermi-Dirac distribution proposed 
in~\cite{turkey1,Pennini1995309,Teweldeberhan:2005wq,Conroy:2008uy,Conroy:2010wt}  
which uses 
\begin{equation}
n^{FD}_T(E) \equiv 
\frac{1}{1+\exp_q\left(\frac{E-\mu}{T}\right)}  .
\label{tsallis-fd}
\end{equation}
where the function $\exp_q(x)$ is defined as
\begin{equation}
\exp_q(x) \equiv \left\{
\begin{array}{l l}
\left[1+(q-1)x\right]^{1/(q-1)}&~~\mathrm{if}~~~x > 0 \\
\left[1+(1-q)x\right]^{1/(1-q)}&~~\mathrm{if}~~~x \leq 0 \\
\end{array} \right.
\label{tsallis-fd1}
\end{equation}
and, in the limit where $q \rightarrow 1$ reduces to the standard exponential:
$$
\lim_{q\rightarrow 1}\exp_q(x)\rightarrow \exp(x)
$$
The form given in Eq.~(\ref{tsallis-fd}) will be 
referred to as the Tsallis-FD distribution. The Bose-Einstein version
(given below) will be referred to as the Tsallis-BE 
distribution~\cite{Chen200265}. 

All forms of the   Tsallis distribution introduce a  
new parameter $q$. In practice this
parameter is always close to 1, e.g. in the 
results obtained by the ALICE and CMS collaborations
typical values for the parameter $q$ can be obtained  
from fits to the transverse momentum distribution for identified 
charged particles ~\cite{Aamodt:2011zj} and are in the range 1.1 to 1.2 (see 
below).
The value of $q$ should thus be considered as never being far from 1, 
deviating from it by  20\% at most. An analysis of the composition of 
final state particles leads to a similar result~\cite{Cleymans:2008mt}
for the parameter $q$.

\begin{figure}
\begin{center}
\includegraphics[width=\textwidth,height=10cm]{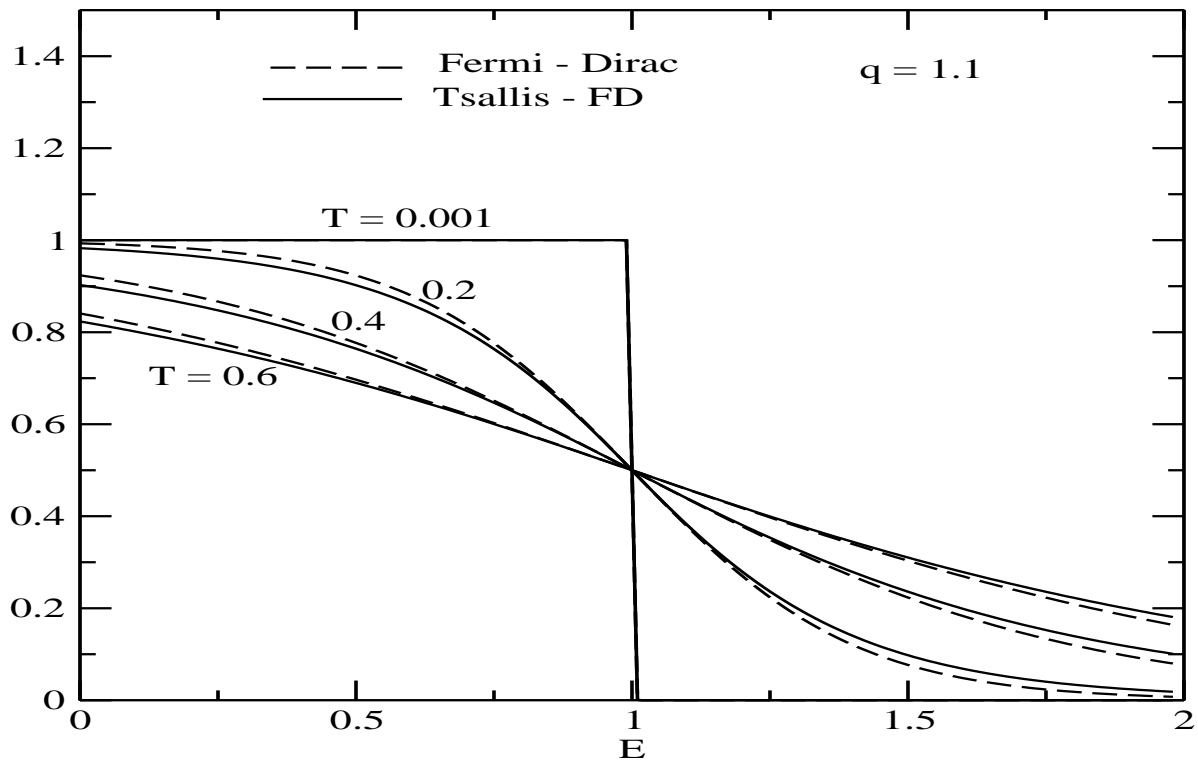}
\caption{Comparison between the Fermi-Dirac and Tsallis-FD distributions 
as a function of the energy $E$, keeping the Tsallis parameter $q$ fixed, for 
various values of the temperature $T$. 
The chemical potential is kept equal to one in all curves, the units
are arbitrary.}
\label{compare-tsallis-fd}
\end{center}
\end{figure}

The classical limit  will be referred to
as Tsallis-B  distribution (the B stands for 
the fact that it reduces to the Boltzmann distribution
 in the limit where $q\rightarrow 1$)
and is given by result~\cite{Tsallis:1987eu,Tsallis:1998ab}
\begin{equation}
n_T^B(E) = \left[ 1 + (q-1) \frac{E-\mu}{T}\right]^{-\frac{1}{q-1}} .
\label{tsallis}
\end{equation}
Note that we do not use the normalized $q$-probabilities which have been 
proposed in Ref.~\cite{Tsallis:1998ab} since we use here mean occupation numbers
which do not need to be normalized.  
In the limit where 
$q\rightarrow 1$ all distributions coincide with
the standard statistical distributions:
\begin{eqnarray}
\lim_{q\rightarrow 1} n_T^B(E)    &=& n^B(E), \\
\lim_{q\rightarrow 1} n_T^{FD}(E) &=& n^{FD}(E), \\
\lim_{q\rightarrow 1} n_T^{BE}(E) &=& n^{BE}(E) .
\end{eqnarray}
A derivation of the Tsallis distribution, based on the Boltzmann equation,
has been given in Ref.~\cite{Biro:2005uv,Barnafoldi}.
Numerically the difference between 
Eq.~(\ref{tsallis-fd}) anfd the Fermi-Dirac distribution is 
small, as shown in Fig.~(\ref{compare-tsallis-fd})
 for a 
value of $q = 1.1$. 

The Tsallis-B distribution is 
always larger than the Boltzmann one if $q>1$. Taking 
into account the large $p_T$ results for 
particle production we will only consider this possibility in this paper.
As a consequence, in order to keep
the particle yields the same, the Tsallis distribution always leads to
smaller values of the freeze-out temperature for the same set of 
particle yields~\cite{Cleymans:2008mt}.
\section{Thermodynamic Consistency}
%
%
The first and second laws of thermodynamics lead to the following two 
differential relations~\cite{deGroot:1980aa}
\begin{eqnarray}
 d\epsilon = Tds + \mu dn,\label{a5}\\
dP = sdT + nd\mu.\label{a51}
\end{eqnarray}
where $\epsilon = E/V$, $s = S/V$ and $n = N/V$ are the energy, entropy and particel
densities respectively.
Thermodynamic consistency 
requires that the following relations be satisfied
\begin{eqnarray}\label{a6}
 T &=& \left.\frac{\partial \epsilon}{\partial s}\right|_{n},\label{a61}\\
 \mu &=&\left.\frac{\partial \epsilon}{\partial n}\right|_{s},\\
 n &=& \left.\frac{\partial P}{\partial \mu}\right|_{T},\label{a63}\\
 s &=& \left.\frac{\partial P}{\partial T}\right|_{\mu}.\label{a64}
\end{eqnarray}
The pressure, energy density and entropy density are all given by corresponding
integrals over Tsallis distributions and the derivatives have 
to reproduce the corresponding physical quantities. 
For completeness, in the next section, we derive
Tsallis thermodynamics using the maximal entropy principle
and discuss  quantum 
$q$-statistics in particular Bose-Einstein and Fermi-Dirac distribution by maximizing the entropy of the system for quantum
 distributions. This extends the derivation of Ref.~\cite{Conroy:2010wt}.
We will show that the consistency conditions given above are indeed 
obeyed by the Tsallis-FD distribution.
%
%
%
\section{Quantum Statistics}
The entropy in standard statistical mechanics for fermions is given 
in the large volume limit by:
\begin{eqnarray}\label{genfermi}
S^{FD}=-gV&&\displaystyle\int \frac{d^3p}{(2\pi)^3}
\left[ n^{FD}\ln n^{FD}\right.\nonumber\\
&&+\left.(1-n^{FD}) \ln(1-n^{FD}) \right],
\end{eqnarray}
where $g$ is the degeneracy factor and $V$ the volume of the system. For 
simplicity Eq.~(\ref{genfermi}) refers to one particle species but can 
be easily generalized to many.
In the limit where momenta are quantized, which is given by:
\begin{equation}\label{genfermi-small}
S^{FD}=-g\displaystyle\sum_{i}\left[ n_i\ln n_i+(1-n_i) \ln(1-n_i) \right],
\end{equation}
For convenience we will work with the discrete form in the rest of this section.
The large volume limit can be recovered with the standard
replacement:
\begin{equation}
\displaystyle\sum_i \rightarrow V\displaystyle\int \frac{d^3p}{(2\pi)^3}
\end{equation}
The generalization, using the Tsallis prescription, leads to  
\cite{turkey1,Pennini1995309,Teweldeberhan:2005wq}
\begin{equation}\label{genfermi-tsallis}
S^{FD}_T=-g\displaystyle\sum_{i}\left[ n_{i}^{q}\ln_{q} n_{i}+(1-n_{i})^{q} \ln_{q}(1-n_{i}) \right],
\end{equation}
where use has been made of the function 
\begin{equation}
\ln_q (x)\equiv \frac{x^{1-q}-1}{1-q} , \label{suba} 
\end{equation}
often referred to as q-logarithm. 
It can be easily shown that in the limit where the Tsallis parameter $q$
tends to 1 one has:
\begin{equation}
\lim_{q\rightarrow 1}\ln_q (x) = \ln (x) . 
\end{equation} 
In a similar vein,  the generalized form of the  entropy 
for bosons is given by
\begin{equation}\label{genboson-tsallis}
S^{BE}_T=- g\displaystyle\sum_i\left[ n_i^q\ln_{q} n_i-(1+n_i)^q \ln_{q}(1+n_i) \right],
\end{equation}
In the limit $q\rightarrow 1$ equations \ref{genfermi-tsallis} and 
\ref{genboson-tsallis} reduce to the standard Fermi-Dirac and 
Bose-Einstein distributions.
Further, as we shall presently explain, the formulation 
of a variational principle in terms of  the above equations allows 
to prove the validity of the general relations of thermodynamics. 
One of the relevant constraints 
is given by the average number of particles,
\begin{equation}\label{n1}
\displaystyle\sum_{i} n_{i}^{q} =  N .
\end{equation}

Likewise, the energy of the system gives a constraint,
\begin{equation}\label{n2}
\displaystyle\sum_{i} n_{i}^{q}E_{i} = E .
\end{equation}
it is necessary to have the power $q$ on the left-hand side as no 
thermodynamic consistency would be achieved without it.
The maximization of the entropic measure 
under the constraints 
Eqs.~(\ref{n1}) and (\ref{n2}) leads
to the variational equation:
\begin{equation}\label{n3}
\frac{\delta}{\delta n_{i}}
\left[ S^{FD}_T+\alpha (N -\displaystyle\sum_i n_i^q) +\beta(E -\displaystyle\sum_i n_i^qE_i)\right] =0,
\end{equation}
where $\alpha$ and $\beta$ are Lagrange multipliers 
associated, respectively, with the total number of particles and 
the total energy. Differentiating each 
expression in iEq.~(\ref{n3})
\begin{equation}\label{n4}
\frac{\delta}{\delta n_{i}}\left(  S^{FD}_T\right) 
=\frac{q}{q-1}\left[\left( \frac{1-n_{i}}{n_{i}}\right)^{q-1}-1 \right] n_{i}^{q-1},
\end{equation}
\begin{equation}\label{n5}
\frac{\delta}{\delta n_{i}}\left( N - \displaystyle\sum_in_i^q\right) 
= -qn_i^{q-1},
\end{equation}
and
\begin{equation}\label{n6}
\frac{\delta}{\delta n_{i}}\left(E - \displaystyle\sum_in_i^qE_i\right)
 =-qE_in_i^{q-1}.
\end{equation}
By substituting Eqs.~(\ref{n4}), (\ref{n5}) and 
(\ref{n6}) into (\ref{n3}), we obtain
\begin{equation}\label{n7}
qn_{i}^{q-1} \left\lbrace \frac{1}{q-1}\left[  -1+\left(\frac{1-n_{i}}{n_{i}} \right)^{q-1} \right]
-\beta E_{i}-\alpha \right\rbrace=0.
\end{equation}
Which can be rewritten as
\begin{equation}\label{n8}
\frac{1}{q-1}\left[  -1+\left(\frac{1-n_{i}}{n_{i}} \right)^{q-1} \right]=\beta E_{i}+\alpha,
\end{equation}
and, by rearranging Eq.~(\ref{n8}), we get
\begin{displaymath}
\frac{1-n_{i}}{n_{i}}=\left[ 1+(q-1)(\beta E_{i}+\alpha)\right]^{\frac{1}{q-1}},
\end{displaymath}
which gives the Tsallis-FD form referred to earlier
in this paper
as~\cite{turkey1,Pennini1995309,Teweldeberhan:2005wq}
\begin{eqnarray}
n_{i}&=&\frac{1}{\left[ 1+(q-1)(\beta E_{i}+\alpha)\right]^{\frac{1}{q-1}}+1},\nonumber \\
     &=&\frac{1}{\left[\exp_{q}(\alpha +\beta E_{i} )\right] +1}.
\end{eqnarray}
Using a similar approach one can also determine the tsallis-BE distribution by 
starting from the extremum of the entropy subject to the same two conditionss:
\begin{equation}\label{n10}
\frac{\delta}{\delta n_{i}}\left[ S^{BE}_T+\alpha(N -\displaystyle\sum_{i} 
n_i^{q}) +\beta(E -\displaystyle\sum_{i} n_{i}^{q}E_{i})\right] =0,
\end{equation}
which leads to
\begin{eqnarray}
n_{i}&=&\frac{1}{\left[ 1+(q-1)(\beta E_{i}+\alpha)\right]^{\frac{1}{q-1}}-1},\nonumber \\
     &=&\frac{1}{\left[\exp_{q}((E_{i} -\mu)/T)\right]-1}  .\label{q}
\end{eqnarray}
where the usual identifications $\alpha = -\mu/T$  and $\beta = 1/T$ have been made. 
\section{Proof of Thermodynamical Consistency}
In order to use the above expressions it has to be shown that they satisfy 
the thermodynamic consistency conditions. To show this in detail 
we use the first law of thermodynamics~\cite{deGroot:1980aa}
\begin{equation}\label{a10}
 P = \frac{-E + TS + \mu N}{V}, 
\end{equation}
and take the partial derivative with respect to $\mu$ in
order to check for thermodynamic consistency, it leads to
\begin{eqnarray}\label{a11}
\left.\frac{\partial P}{\partial \mu}\right|_T & = &
\frac{1}{V}
\left[-\frac{\partial E}{\partial \mu} +T\frac{\partial S}{\partial \mu} + N + \mu\frac{\partial N}{\partial \mu}\right],\nonumber \\  
& = &\frac{1}{V}\left[N + \displaystyle\sum_{i}-\frac{T}{q-1}\left(1 + 
(q-1)\frac{E_{i} -\mu}{T}\right)\frac{\partial n_{i}^{q}}{\partial \mu} \right.\nonumber \\
&& \left.+\frac{Tq(1-n_{i})^{q-1}}{q-1}\frac{\partial n_{i}}{\partial \mu}
\right],
\end{eqnarray}
then, by explicit calculation
\begin{displaymath}
 \frac{\partial n_{i}^{q}}{\partial \mu} = \frac{qn_{i}^{q+1}}{T}\left[1+(q-1)\frac{E_{i}-\mu}{T}\right]^{-1+\frac{1}{1-q}},
\end{displaymath}
\begin{displaymath}
 \frac{\partial n_{i}}{\partial \mu} = \frac{n_{i}^{2}}{T}\left[1+(q-1)\frac{E_{i}-\mu}{T}\right]^{-1+\frac{1}{1-q}},
\end{displaymath}
and
\begin{displaymath}
\left(1-n_{i}\right)^{q-1} = n_{i}^{q-1}\left[1+\frac{(q-1)(E_{i}-\mu)}{T}\right].
\end{displaymath}
Introducing this  into equation \ref{a11},  yields
\begin{equation}\label{a12}
\left. \frac{\partial P}{\partial \mu}\right|_T = n,
\end{equation}
which proves the thermodynamical consistency \ref{a63}. 

We also calculate explicitly the relation in equation \ref{a61} can be rewritten as
\begin{eqnarray}\label{a13}
\left. \frac{\partial E}{\partial S}\right|_n& = &\frac{\frac{\partial E}{\partial T}dT + \frac{\partial E}{\partial \mu}d\mu}{\frac{\partial S}{\partial T}dT
 + \frac{\partial S}{\partial \mu}d\mu},\nonumber \\
& = &\frac{\frac{\partial E}{\partial T} + \frac{\partial E}{\partial \mu}\frac{d\mu}{dT}}{\frac{\partial S}{\partial T}
 + \frac{\partial S}{\partial \mu}\frac{d\mu}{dT}},
\end{eqnarray}
since $n$ is kept fixed one has the additional constraint
\begin{displaymath}
 dn = \frac{\partial n}{\partial T}dT + \frac{\partial n}{\partial \mu}d\mu = 0,
\end{displaymath}
leading to
\begin{equation}\label{a18}
 \frac{d\mu}{dT} = -\frac{\frac{\partial n}{\partial T}}{\frac{\partial n}{\partial \mu}}.
\end{equation}
Now, we rewrite \ref{a13} and \ref{a18} in terms of the following 
expressions
\begin{displaymath}
 \frac{\partial E}{\partial T} = \displaystyle\sum_{i} qE_{i}n_{i}^{q-1}\frac{\partial n_{i}}{\partial T}, 
\end{displaymath}
\begin{displaymath}
 \frac{\partial E}{\partial \mu} = \displaystyle\sum_{i} qE_{i}n_{i}^{q-1}\frac{\partial n_{i}}{\partial \mu}, 
\end{displaymath}
\begin{displaymath}
 \frac{\partial S}{\partial T} = \displaystyle\sum_{i} q\left[\frac{-n_{i}^{q-1}+(1-n_{i})^{q-1}}{q-1}\right]\frac{\partial n_{i}}{\partial T}, 
\end{displaymath}
\begin{displaymath}
 \frac{\partial S}{\partial \mu} = \displaystyle\sum_{i}q\left[\frac{-n_{i}^{q-1}+(1-n_{i})^{q-1}}{q-1}\right]\frac{\partial n_{i}}{\partial \mu}, 
\end{displaymath}
\begin{displaymath}
 \frac{\partial n}{\partial T} = \frac{1}{V}\left[\displaystyle\sum_{i} qn_{i}^{q-1}\frac{\partial n_{i}}{\partial T}\right], 
\end{displaymath}
and
\begin{displaymath}
 \frac{\partial n}{\partial \mu} = \frac{1}{V}\left[\displaystyle\sum_{i} qn_{i}^{q-1}\frac{\partial n_{i}}{\partial \mu}\right]. 
\end{displaymath}
By introducing the above relations into equation \ref{a13}, the numerator 
 of equation \ref{a13} becomes
\begin{eqnarray}\label{a19}
\frac{\partial E}{\partial T} &+& \frac{\partial E}{\partial \mu}\frac{d\mu}{dT} 
 = \displaystyle\sum_{i} qE_{i}n_{i}^{q-1}\frac{\partial n_i}{\partial T}\nonumber \\
&&-\frac{\displaystyle\sum_{i,j} q^{2}E_{j}\left(n_{i}n_{j}\right)^{q-1}\frac{\partial n_{j}}{\partial \mu}\frac{\partial n_{i}}{\partial T}}{\displaystyle\sum_{j} 
qn_{j}^{q-1}\frac{\partial n_{j}}{\partial \mu}},\nonumber \\ 
& =& \frac{\displaystyle\sum_{i,j} qE_{i}\left(n_{i}n_{j}\right)^{q-1}C_{ij}}
{\displaystyle\sum_{j} 
n_{j}^{q-1}\frac{\partial n_{j}}{\partial \mu}}.
\end{eqnarray}
Where the abbreviation 
\begin{equation}
 C_{ij}\equiv \left(n_{i}n_{j}\right)^{q-1}\left[\frac{\partial n_{i}}{\partial T}\frac{\partial n_{j}}{\partial \mu}-
 \frac{\partial n_{j}}{\partial T}\frac{\partial n_{i}}{\partial \mu}\right],
\end{equation}
has been introduced. 
One can rewrite the denominator part of equation \ref{a13}  as
\begin{eqnarray}\label{a20}
\frac{\partial S}{\partial T} + \frac{\partial S}{\partial \mu}\frac{d\mu}{dT} & = 
 \frac{\displaystyle q\sum_{i,j}\left[{-n_{i}^{q-1}+(1-n_{i})^{q-1}}\right]n_{j}^{q-1}
 C_{i,j}}
 {(q-1)\displaystyle\sum_{j} 
n_{j}^{q-1}\frac{\partial n_{j}}{\partial \mu}},\nonumber \\
& = \frac{\displaystyle q\sum_{i,j}(E_{i}-\mu)\left(n_{i}n_{j}\right)^{q-1}
 C_{i,j}}
 {T\displaystyle \sum_{j} 
n_{j}^{q-1}\frac{\partial n_{j}}{\partial \mu}},
\end{eqnarray}
where
\begin{displaymath}
 \frac{-n_{i}^{q-1}+(1-n_{i})^{q-1}}{q-1} = \frac{(E_{i}-\mu)}{T}n_{i}^{q-1},
\end{displaymath}
hence, by substituting equation \ref{a19} and \ref{a20} in to \ref{a13}, we find
\begin{equation}\label{a21}
\left. \frac{\partial E}{\partial S}\right|_n = T\frac{\displaystyle\sum_{i,j}E_{i}C_{ij}}
{\displaystyle\sum_{i,j}(E_{i}-\mu)C_{ij}},
\end{equation}
since $\displaystyle\sum_{i,j} C_{ij} = 0$, this finally leads to the desired result
\begin{equation}
\left.\frac{\partial E}{\partial S}\right|_n = T.
\end{equation}
Hence thermodynamic consistency is satisfied.
 
It has thus been shown that the definitions of temperature and
pressure within the Tsallis formalism 
for non-extensive thermostatistics lead to expressions which
satisfy consistency with the first and second laws of thermodynamics.

%
%
%
\section{Tsallis Fit Details} 
The total number of particles is given by the  integral version of Eq.~(\ref{n1}), 
\begin{equation}
N = gV\displaystyle\int \frac{d^3p}{(2\pi)^3}
\left[1+(q-1)\frac{E-\mu}{T}\right]^{q/(1-q)}  .
\end{equation}
The corresponding (invariant)  momentum distribution deduced from the 
equation above is given by 
\begin{equation}
E\frac{dN}{d^3p} = gVE\frac{1}{(2\pi)^3}
\left[1+(q-1)\frac{E-\mu}{T}\right]^{q/(1-q)},
\end{equation}
which, in terms of the rapidity and transverse mass variables, becomes

\begin{eqnarray}
\frac{dN}{dy\, p_Tdp_T} &=& gV\frac{m_T\cosh y}{(2\pi)^2}\nonumber\\
&&\times\left[1+(q-1)\frac{m_T\cosh y -\mu}{T}\right]^{q/(1-q)},
\end{eqnarray}
At mid-rapidity $y=0$ and for zero chemical potential this reduces to
the following expression
\begin{equation}\label{alice}
\left.\frac{dN}{dp_T~dy}\right|_{y=0} = gV\frac{p_Tm_T}{(2\pi)^2}
\left[1+(q-1)\frac{m_T}{T}\right]^{q/(1-q)}.
\label{final}
\end{equation}
Fits using the above expressions based on the Tsallis-B distribution
 to experimental 
measurements published by the CMS collaboration~\cite{Khachatryan:2011tm}
are shown in Figs.~(2), (3) and (4) are comparable with those 
shown by the CMS  collaboration. Fits to the results 
obtained by the ALICE collaboration~\cite{Aamodt:2011zj} are shown 
in Fig.~(5) for $\pi^-, K^-,\bar{p}$.
The resulting parameters are considerably different 
from those obtained from Eq.~\ref{ALICE-CMS} and are collected 
in Table I.  The most striking feature is that the values of 
the parameter $q$ are fairly stable in the range 1.1 to 1.2 
for all particles 
considered at 0.9 TeV. The  
temperature $T$ cannot be determined very accurately for all hadrons 
but they are consistent with a value around 70 MeV.\\
%
%
%
\begin{table}[ht]
\begin{center}
\begin{tabular}{|c|c|c|}
\hline
Particle & $q$                &$T$ (GeV)      \\
\hline 
$\pi^+$  & 1.154  $\pm$0.036 & 0.0682 $\pm $0.0026 \\ 
$\pi^-$  & 1.146  $\pm$0.036 & 0.0704 $\pm$ 0.0027 \\
$K^+$    & 1.158  $\pm$0.142 & 0.0690 $\pm $0.0223  \\
$K^-$    & 1.157  $\pm$0.139 & 0.0681 $\pm$ 0.0217  \\
$K^0_S$  & 1.134  $\pm$0.079 & 0.0923 $\pm $0.0139  \\
$p$      & 1.107  $\pm$0.147 & 0.0730 $\pm$ 0.0425   \\
$\bar{p}$& 1.106  $\pm$0.158 & 0.0764 $\pm $0.0464  \\
$\Lambda$& 1.114  $\pm$0.047 & 0.0698 $\pm$ 0.0148  \\
$\Xi^-$  & 1.110  $\pm$0.218 & 0.0440 $\pm$ 0.0752  \\
\hline  
\end{tabular}
\caption{Fitted values of the $T$ and $q$ parameters for strange particles 
measured by
the ALICE~\cite{Aamodt:2011zj} and CMS collaborations~\cite{Khachatryan:2011tm} 
using the Tsallis-B form for the momentum distribution. 
}
\end{center}
\end{table}
For clarity we show these results also in Fig.~(\ref{q_value}) for 
the values of 
the parameter $q$ and in Fig.~(\ref{T_value})  for the values
 of the Tsallis parameter $T$.
The striking feature is that the values of $q$ are consistently between 1.1 and 1.2
for all species of hadrons.
\section{Conclusions}
In this paper we have presented a detailed derivation  of the quantum form
of the Tsallis distribution and considered in detail 
the thermodynamic consistency of the resulting distribution.
It was emphasized that an additional power of $q$ is needed to achieve
consistency with the laws of thermodynamics~\cite{Conroy:2010wt}.
The resulting distribution, called Tsallis-B, was compared with 
recent measurements from the ALICE~\cite{Aamodt:2011zj} 
and CMS collaborations~\cite{Khachatryan:2011tm}
and good agreement was obtained. The resulting parameter $q$ which is
a measure for the deviation from a standard Boltzmann distribution was found
to be around 1.11.  The resulting values of the temperature are
also consistent within the  errors and lead to a value of 
around 70 MeV.

\begin{figure}
\begin{center}
\includegraphics[width=\textwidth,height=10cm]{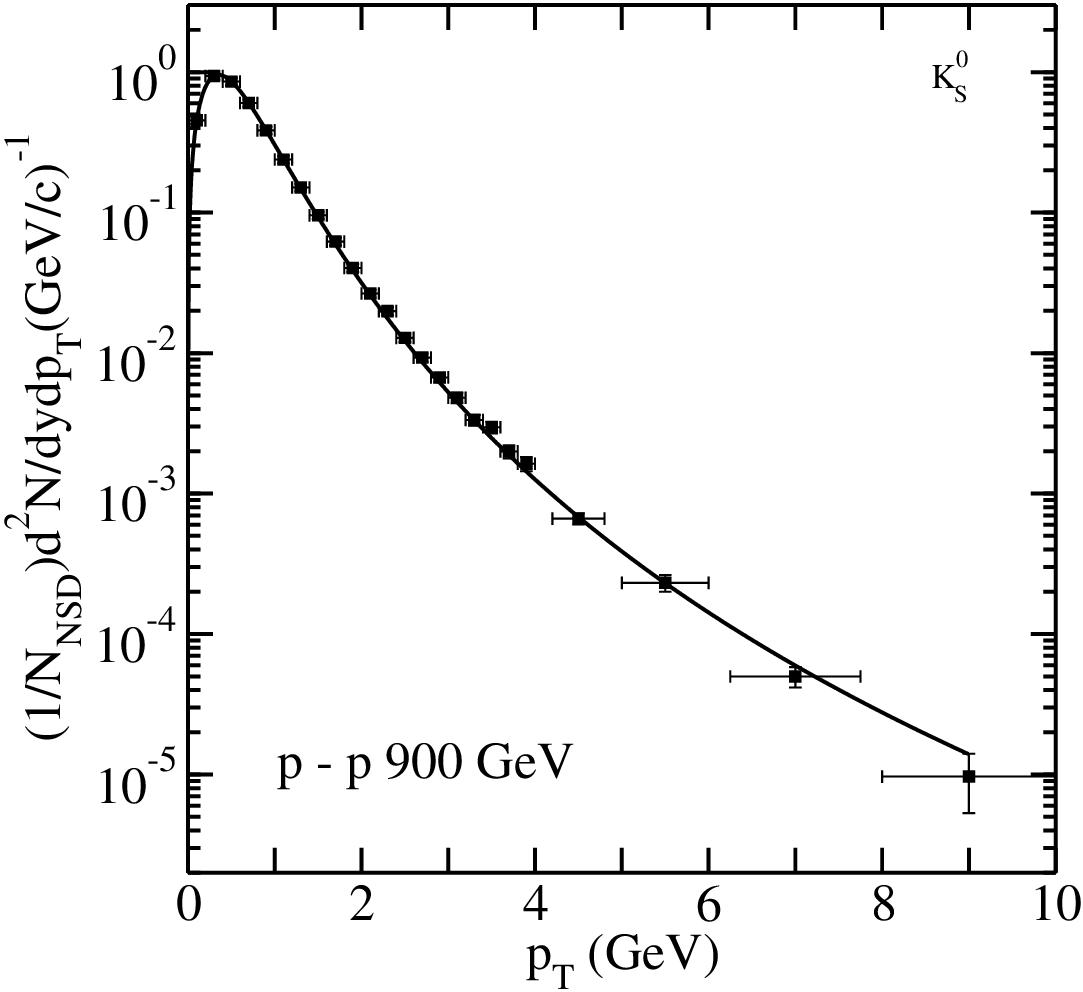}
\caption{Comparison between the 
measured transverse momentum distribution for $K^0_S$ as measured by 
the CMS collaboration~\cite{Khachatryan:2011tm} and  the Tsallis-B distribution.
The full line is a fit using the parameterization
given in Eq.~(\ref{final}) 
to the 0.9 TeV data with the parameters listed in Table I.  
}
\label{CMS-K0S}
\end{center}
\end{figure}
\begin{figure}
\begin{center}
\includegraphics[width=\textwidth,height=10cm]{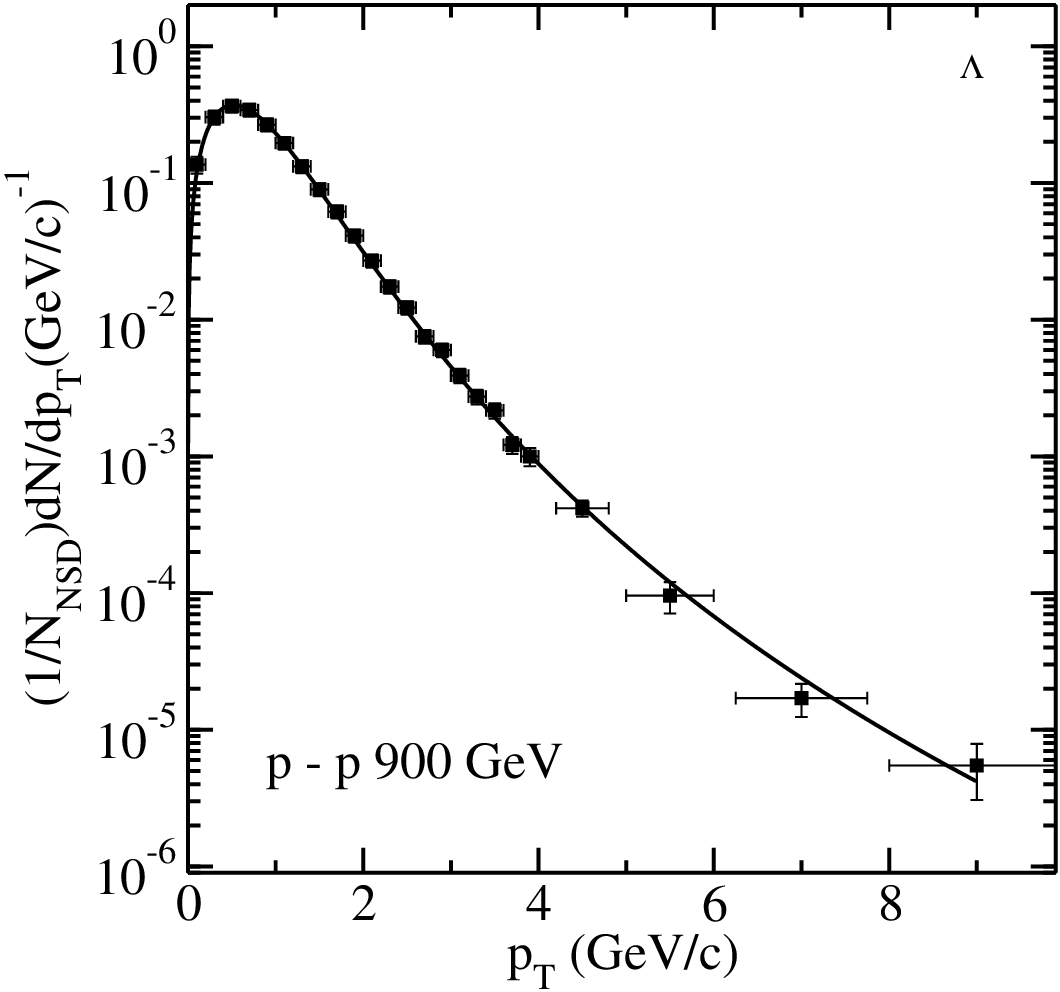}
\caption{Comparison between the 
measured transverse momentum distribution for $\Lambda$ as measured by 
the CMS collaboration~\cite{Khachatryan:2011tm} and  the Tsallis-B distribution.
The full line is a fit using the parameterization
given in Eq.~(\ref{final}) to the 0.9 TeV data with the parameters listed in Table I.  
}
\label{CMS-Lambda}
\end{center}
\end{figure}
\begin{figure}
\begin{center}
\includegraphics[width=\textwidth,height=10cm]{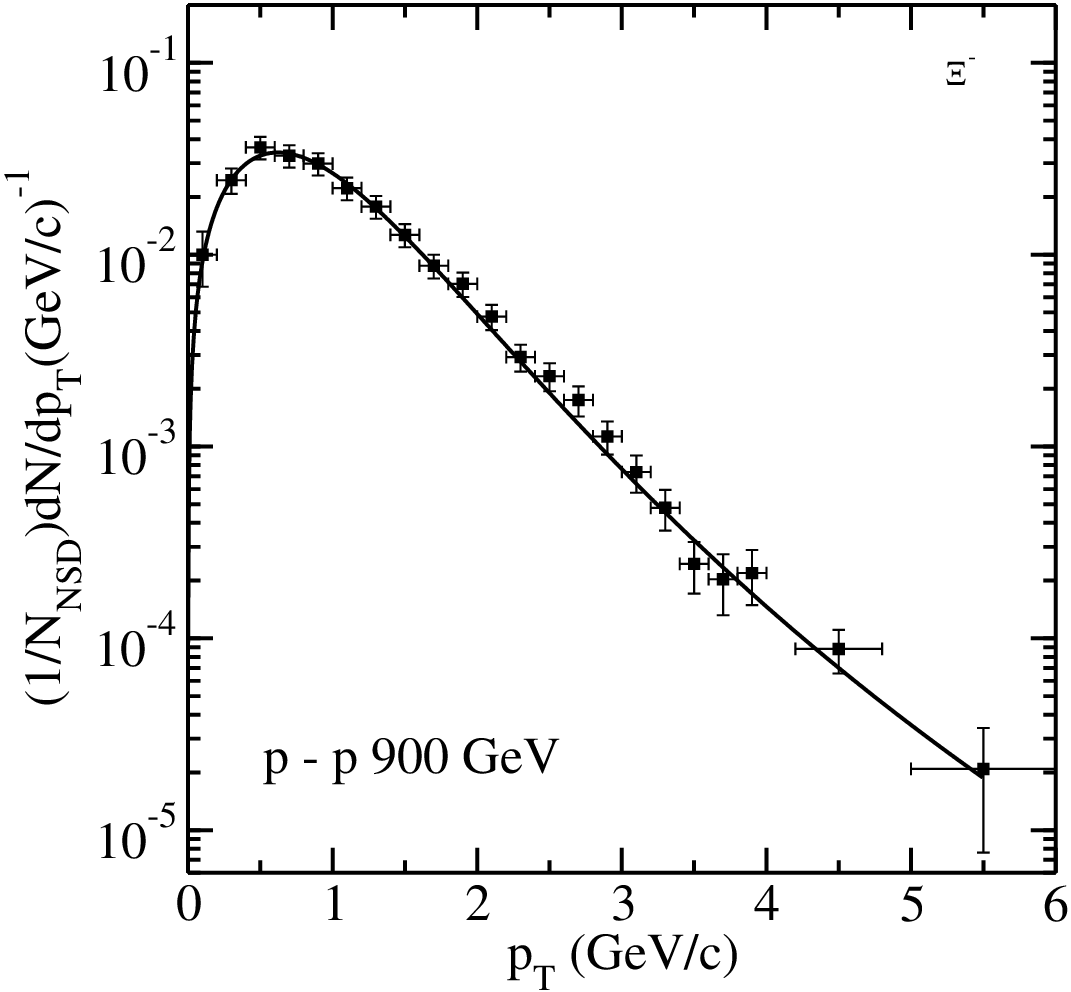}
\caption{Comparison between the 
measured transverse momentum distribution for $\Xi^-$ as measured by 
the CMS collaboration~\cite{Khachatryan:2011tm} and  the Tsallis-B distribution.
The full line is a fit using the parameterization given in Eq.~(\ref{final})
 to the 
 0.9 TeV data with the parameters listed in Table I.  
 }
\label{CMS-Xi}
\end{center}
\end{figure}
\begin{figure}
\begin{center}
\includegraphics[width=\textwidth,height=10cm]{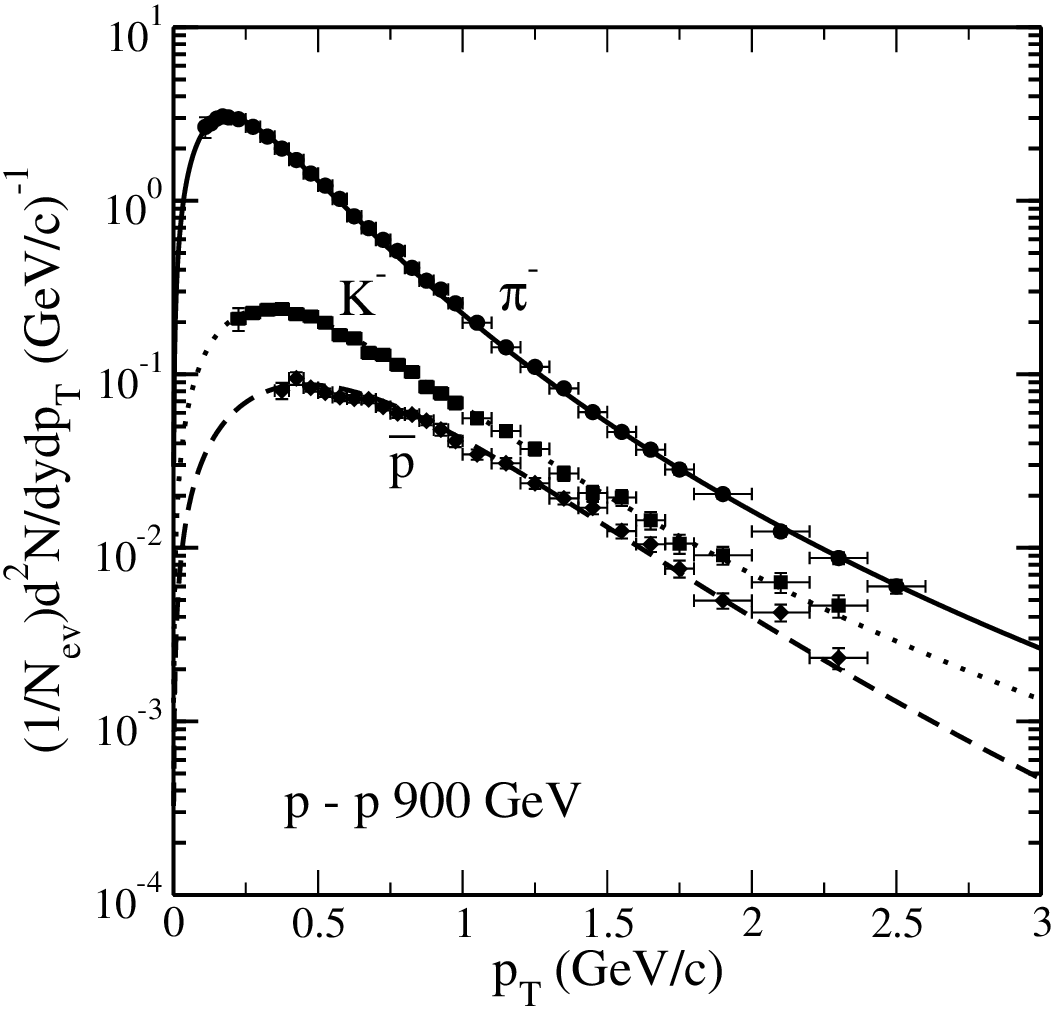}
\caption{Comparison between the 
measured transverse momentum distribution for $\pi^-$, $K^-$ and $\bar{p}$
 as measured by 
the ALICE collaboration~\cite{Aamodt:2011zj} and  the Tsallis-B distribution.
The lines are fits using the parameterization given in Eq.~(\ref{final}) 
to the 0.9 TeV data with the parameters listed in Table I. Full line is for $\pi^-$, 
the dotted line is for $K^-$, the dashed line is for anti-protons.}
\label{ALICE}
\end{center}
\end{figure}
\begin{figure}
\begin{center}
\includegraphics[width=\textwidth,height=10cm]{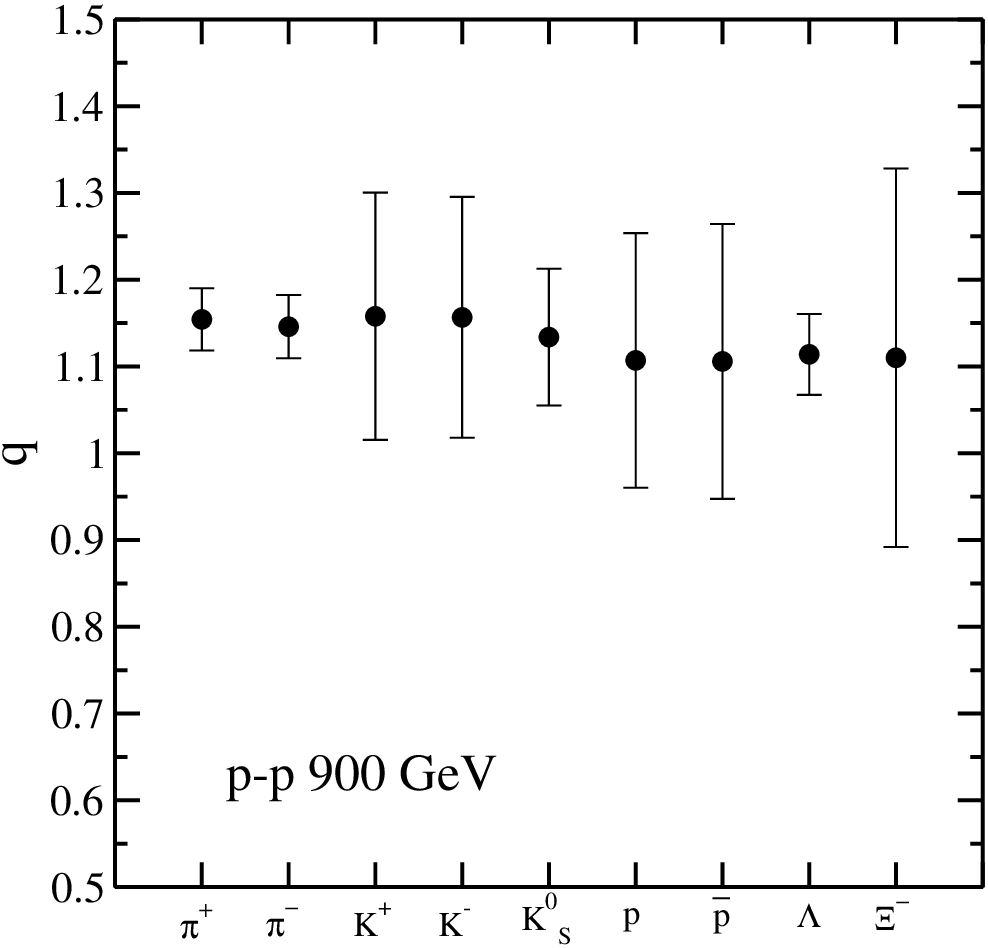}
\caption{Values of the Tsallis parameter $q$ for different  species of hadrons. }
\label{q_value}
\end{center}
\end{figure}
\begin{figure}
\begin{center}
\includegraphics[width=\textwidth,height=10cm]{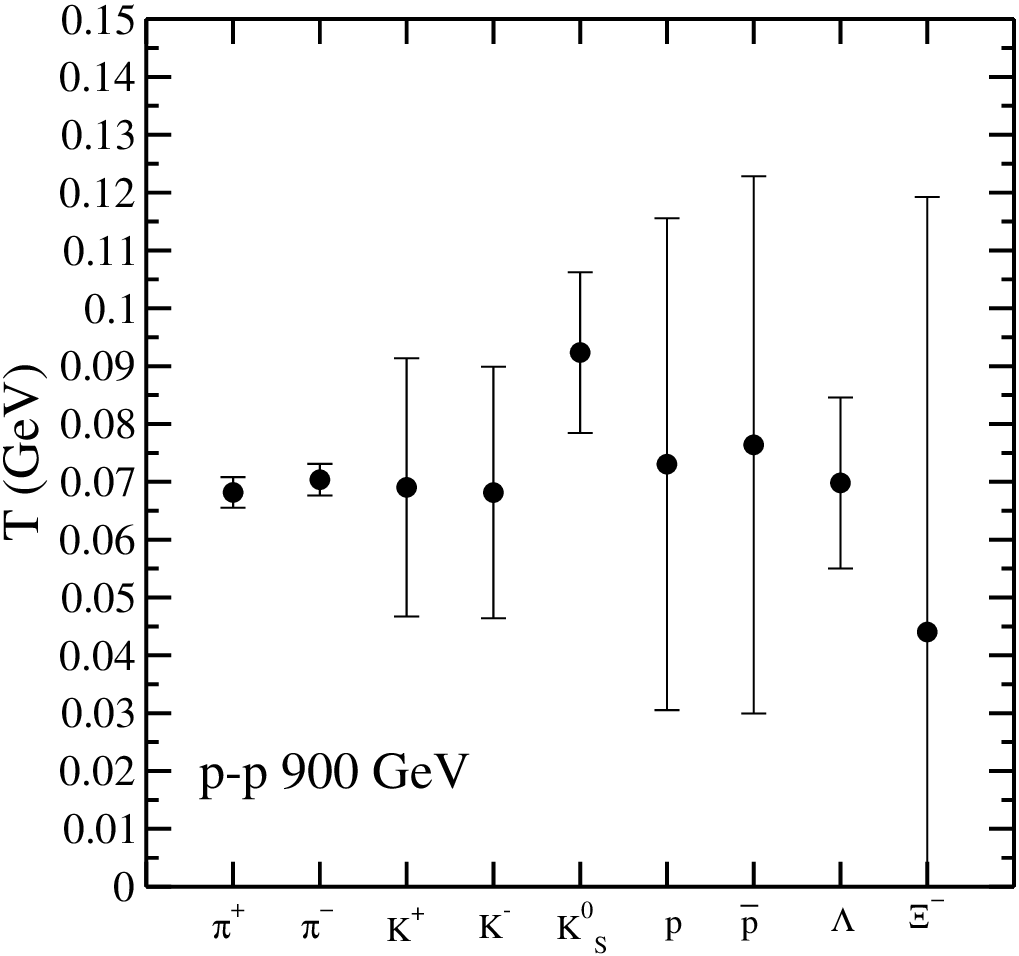}
\caption{Values of the Tsallis temperature $T$ for different species of hadrons. }
\label{T_value}
\end{center}
\end{figure}
\end{document}